\documentclass[a4paper,11pt]{article}
\usepackage{pos}
\usepackage{epsfig}
\usepackage{amsfonts}
\usepackage{amsmath}
\usepackage{bbm}
\usepackage{multirow}
\usepackage[colorinlistoftodos]{todonotes}

\newcommand{\be}{\begin{equation}}
\newcommand{\ee}{\end{equation}}
\newcommand{\ba}{\begin{eqnarray}}
\newcommand{\ea}{\end{eqnarray}}
\newcommand{\rmi}[1]{{\mbox{\scriptsize #1}}}
\newcommand{\tab}{Tab.~}
\newcommand{\fig}{Fig.~}

\newcommand{\eq}{Eq.~}

\newcommand{\VT}{V^{ }_\rmi{$T$}}
\newcommand{\alphas}{\alpha_\rmi{s}}
\newcommand{\mD}{m_\rmi{D}}
\newcommand{\rmO}{{\mathcal{O}}}
\newcommand{\bmu}{\bar\mu}
\newcommand{\Nf}{N_{\rm f}}
\newcommand{\Nc}{N_{\rm c}}

\newcommand{\CA}{\Nc}
\newcommand{\CF}{C_\rmi{F}}
\newcommand{\TF}{T_\rmi{F}}

\newcommand{\nn}{\nonumber \\}

\title{Study of charm and beauty in QGP from unquenched lattice QCD}

\author*[a,b]{Sajid Ali}
\author[a]{Dibyendu Bala}
\author[a]{Olaf Kaczmarek}
\author[c]{Hai-Tao Shu}
\author[a]{Tristan Ueding}

\affiliation[a]{University of Bielefeld, Faculty of Physics, Universit\"atsstr.~25, 
D-33615 Bielefeld, Germany}
\affiliation[b]{Government College University Lahore, Department of Physics, Lahore
54000, Pakistan}
\affiliation[c]{Institut f\"ur Theoretische Physik, Universit\"at Regensburg, D-93040
Regensburg, Germany}

\emailAdd{sajid.ali@physik.uni-bielefeld.de}

\abstract{We present charmonium and bottomonium correlators and corresponding 
reconstructed spectral functions from full QCD calculations in the pseudoscalar 
channel. Correlators are obtained using a mixed-action approach, clover-improved 
Wilson valence quarks on gauge field configurations generated with $N_f=2+1$ HISQ sea 
quarks, with physical strange quark masses and light quark masses corresponding to $m_\pi=315$~MeV.
The charm and bottom quark masses are tuned to reproduce 
the experimental mass spectrum of the spin averaged quarkonium vector mesons 
from the particle data group
\cite{ParticleDataGroup:2022pth}.
For the spectral 
reconstruction, we use models based on perturbative spectral functions from different 
frequency regions like resummed thermal contributions around the threshold from pNRQCD
\cite{Laine:2007gj} and vacuum contributions well above the threshold
\cite{Burnier:2012ts}. We show preliminary results of the reconstructed spectral 
function obtained for the first time in our study for full QCD.}
\FullConference{%
  The 39th International Symposium on Lattice Field Theory (Lattice2022),\\
  8-13 August, 2022 \\
  Bonn, Germany 
}
\begin{document}
\maketitle
\section{Introduction}
The spectrum of bound states of heavy quark and anti-quark pairs, the so-called 
quarkonia may receive thermal modifications in the hot QCD medium. To investigate such 
thermal effects the correlator in the pseudoscalar channel is an ideal choice because, 
unlike the vector channel, the spectral function embedded in it has no transport peak 
in the low-frequency region i.e. $|\omega|<<T^2/M$
\cite{tp,umeda, Karsch:2003wy,Aarts:2005hg}, where $M$ is a heavy quark mass, and $T$ 
is the temperature. The Consideration of the pseudoscalar channel excludes the modeling 
of the transport peak. The Euclidean time pseudoscalar correlator reads
\begin{align}
 G_\rmi{PS}(\tau) & \equiv M_{B}^2 \int_{\vec{x}} 
 \Bigl\langle (\bar\psi \gamma^{ }_5 \psi) (\tau,\vec{x})
 (\bar\psi \gamma^{ }_5  \psi) (0,\vec{0}) \Bigr\rangle_c,
\end{align}
where $M_B$ is the bare quark mass and $\psi$ is the Dirac spinor.  
$\langle...\rangle^{ }_{c}$ represents that only the connected diagrams are considered in this study.
The correlator $G_{PS}(\tau)$ is computed numerically on the 4d hypercubic lattice and 
it is related to the corresponding spectral function $\rho_\rmi{PS}(\omega)$ through 
the equation
\be
  G^{ }_\rmi{PS}(\tau) \; = \; 
 \int_0^\infty
 \frac{{\rm d}\omega}{\pi} \rho^{ }_\rmi{PS} (\omega)
 \frac{\cosh \left( \left(\frac{1}{2T} - \tau\right)\omega \right) }
 {\sinh\left( \frac{\omega}{2 T} \right) } 
 \;.\label{eq-corr-spec}
\ee
For given $G^{ }_\rmi{PS}(\tau)$,~\eq(\ref{eq-corr-spec}) can be solved for 
$\rho^{ }_\rmi{PS}(\omega)$ but the solution is not unique
\cite{Kaczmarek:2022ffn,Rothkopf:2019ipj}. The nature of the function 
$\frac{\cosh \left( \left(\frac{1}{2T} - \tau\right)\omega \right) } {\sinh\left( \frac{\omega}{2 T} \right) } $
is such that a small change in $G^{ }_\rmi{PS}(\tau)$ may lead to large uncertainty in 
the spectral function. There might be several spectral functions that all fit into the
correlator data, but which one is correct is hard to determine. Therefore we need some 
theoretically or phenomenologically motivated input to constrain the search space. 
One of the possibilities is to construct the spectral function by taking input from 
perturbation theory or effective field theories like pNRQCD.
In this study we follow the strategy developed in the quenched approximation \cite{Burnier:2017bod}. 

\section{Lattice Setup}
We compute quarkonium correlators using clover-improved Wilson fermions as valence 
quarks at different temperatures listed in \tab\ref{tab-latticesetup}. 
The 
measurements are done on $N_f$=2+1, $m_l=m_s/5$ HISQ gaugefield configurations generated by the 
HotQCD collaboration. 
The clover 
improvement in the fermion valence action reduces $O(a)$ cut-off effects. This improvement is 
made by choosing the Sheikholeslami-Wohlert coefficient $c_{SW}=\frac{1}{u^3_0}$
\cite{Sheikholeslami:1985ij}, where $u_0$ is the tadpole factor and is calculated as 
the fourth root of the plaquette expectation value calculated on HISQ lattices. 
\begin{table}[!b]
\centering
\begin{tabular}{|c|c|c|c|c|c|c|}
\hline
$\beta$ & $a$[fm] & $a^{-1}$[GeV] & $N_\sigma$ & $N_\tau$ & $T$[MeV] & $\#$ confs\\ \hline
\multirow{4}{*}{8.249} & \multirow{4}{*}{0.028} & \multirow{4}{*}{7.033} & 64 & 64 & 110 & 112\\
&& & 96 & 56 & 126 & 200\\ 
&& & 96 & 32 & 220 & 1703\\  
&& & 96 & 28 & 251 & 621\\ \hline  
\end{tabular}
\caption{Lattice parameters for $N_f$=2+1, $m_l=m_s/5$ HISQ configurations. At 
$\beta$=8.249 the lattice spacing $a$ is obtained using $f_k$-scale parametrization
\cite{Bazavov:2014eq}.} 
\label{tab-latticesetup}
\end{table}
\section{Clover mass tuning on mixed action}
\begin{figure}[hbt!]
\centering
\includegraphics[width=0.49\textwidth]{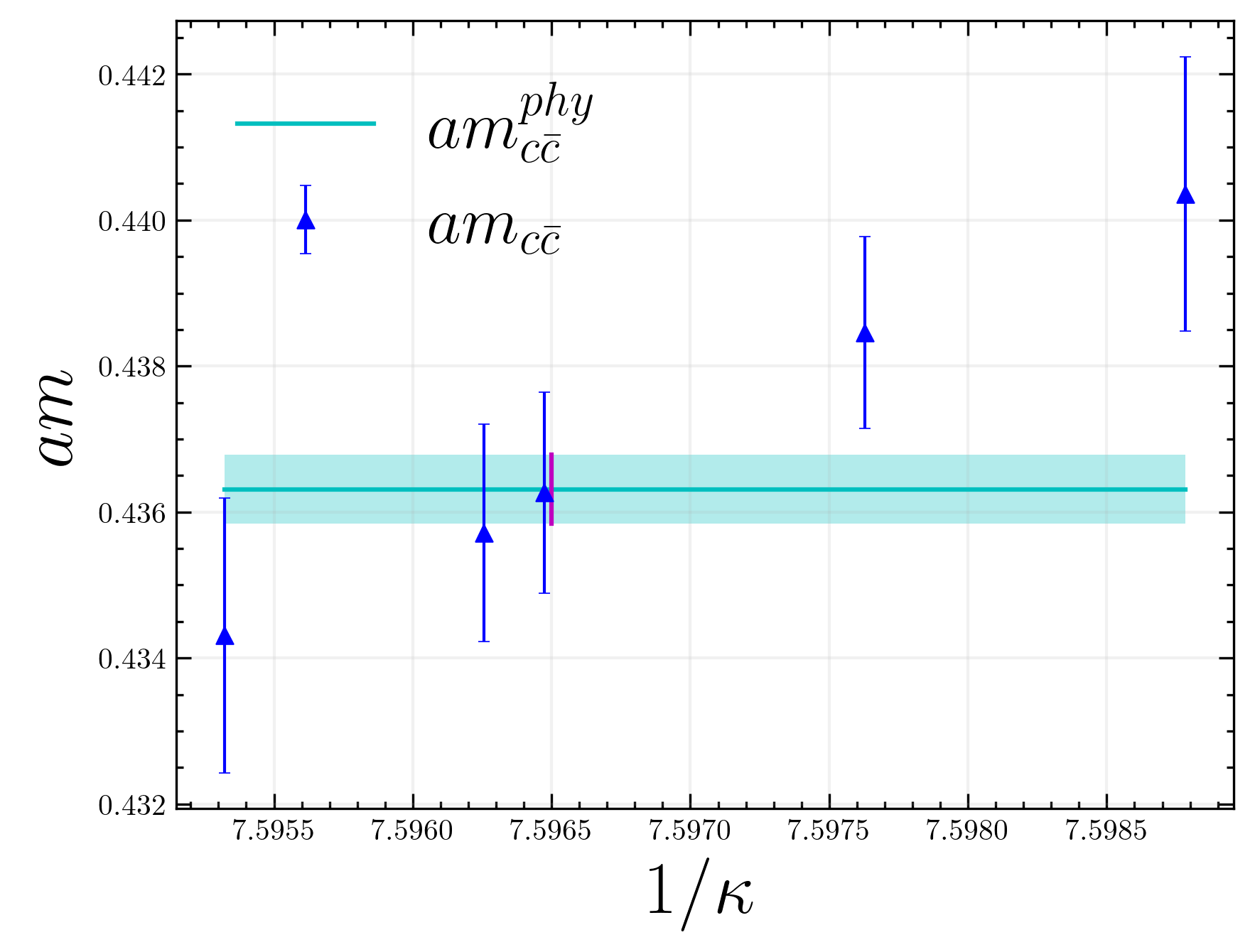}
\caption{Charm quark mass tuning on the mixed action. The solid line represents the 
experimental value of the spin-averaged charmonium in lattice units taken from the PDG. 
Whereas triangles represent masses from lattice calculations for different values of 
$\kappa$ in lattice units.}
\label{tuning}
\end{figure}

The masses of hadrons from lattice QCD also depend on another parameter, the bare quark mass, $m_q$, which for Wilson fermions is tuned by the so-called 
hopping parameter $\kappa$ defined as $\kappa=\frac{1}{8+2m_q}$. To obtain numerical data of mesonic correlators at physical quark 
mass, one needs to tune the quark mass such that the meson mass from the lattice 
is consistent with the corresponding experimental value. In the heavy quark 
sector, we tune to the spin averaged charmonium 
$m_{c\bar{c}}=(m_{\eta_{c}}+3m_{J/\psi})/4$ and bottomonium $m_{b\bar{b}}=(m_{\eta_{b}}+3m_{\Upsilon})/4$. \fig\ref{tuning} illustrates the 
charm quark mass tuning, where $am^{phy}_{c\bar{c}}$ is the physical mass in lattice 
units. The tuned $\kappa$ values for charm and bottom are 0.13164 and 0.11684, 
respectively.
\section{Correlators}
\begin{figure}[hbt!]
\centering
\includegraphics[width=0.49\textwidth]{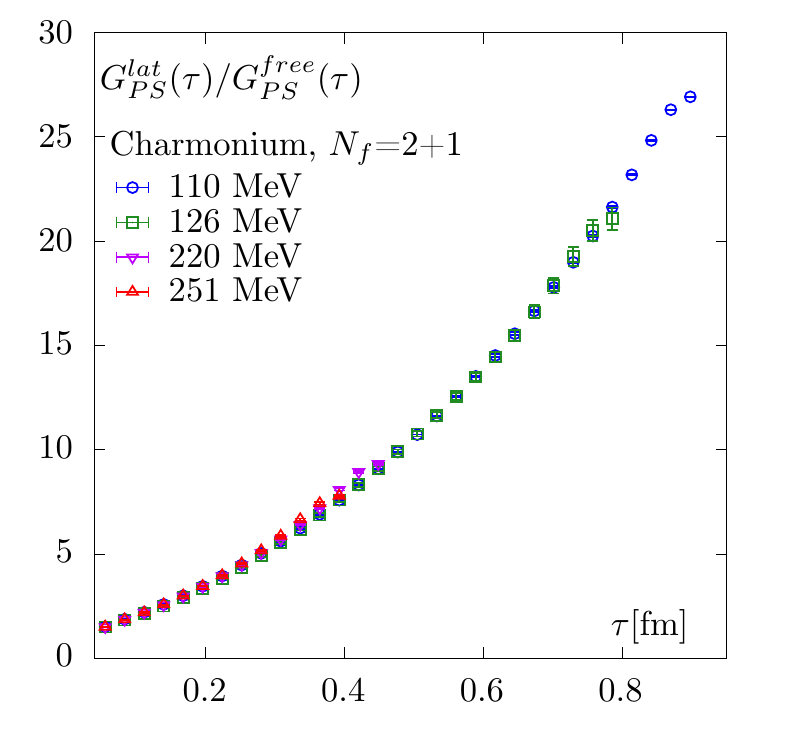}
\includegraphics[width=0.49\textwidth]{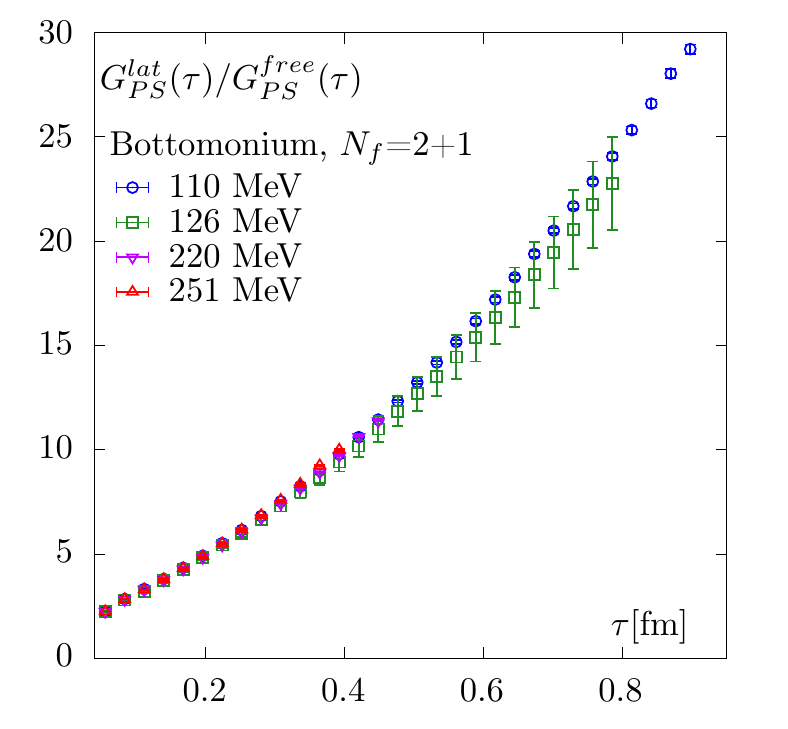}
\caption{Average values of correlator ratios at different temperatures for 
pseudoscalar quarkonium correlators.}
\label{fig-corr}
\end{figure}

To obtain quarkonium spectral functions in the pseudoscalar channel one needs the 
numerical correlator data computed on the lattice. \fig\ref{fig-corr} shows the data 
measured by clover-improved Wilson fermions on large $N_f$=2+1 HISQ gaugefield configurations. 
Instead of the correlator itself, we used the ratio 
$G_\rmi{PS}(\tau)/G^\rmi{free}_\rmi{PS}(\tau)$ so that one can see the behavior as a 
function of $\tau$ at various temperatures, where $G^\rmi{free}_\rmi{PS}(\tau)$ reads \cite{Burnier:2017bod} 
\ba
 \frac{ G_\rmi{PS}^\rmi{free}(\tau) }{ m^2(\bmu^{ }_\rmi{ref}) } 
 & \equiv &   
 \int_{2 M^{ }_\rmi{1S}}^{\infty}
 \! \frac{{\rm d}\omega}{\pi} \, 
 \biggl\{ 
 \frac{\Nc \omega^2}{8\pi}
 \tanh\Bigl( \frac{\omega}{4T} \Bigr)  
 \sqrt{1 - \frac{4 M_\rmi{1S}^2}{\omega^2}} 
 \biggr\}
 \, 
 \frac{\cosh \left(\left(\frac{1}{2T} - \tau\right)\omega\right)}
 {\sinh\left(\frac{\omega}{2T}\right)} 
 \;, \label{GPSfree}
\ea
where $m(\bar{\mu}_\rmi{ref})$ is the running mass at reference scale 
$\bar{\mu}_\rmi{ref}$=2 GeV. We choose the values of $M_\rmi{1S}$ to be 1.5 GeV for charmonium and 
4.7 GeV for bottomonium. One can see that the charmonium 
suffers more thermal effects than the bottomonium as the charmonium correlators at 
high temperatures deviate more from the low-temperature ones. 
%
\section{Perturbative spectral functions in PS-channel}
To construct the quarkonium spectral functions in the pseudoscalar channel
we follow the strategy developed in the quenched approximation \cite{Burnier:2017bod}.
We first
consider thermal contributions around the threshold i.e. $\omega \approx 2 M$. Well 
above the threshold the thermal effects are power suppressed, and the spectral 
function can be replaced by the vacuum one. In the heavy quark mass limit, 
relativistic effects can be ignored. In this case, one can get the pseudoscalar 
spectral function from the vector channel as
\be
 \rho^\rmi{pNRQCD}_\rmi{PS} = 
 \frac{M^2}{3}
 \rho^\rmi{pNRQCD}_\rmi{V} 
 \;, \quad
 \omega \approx 2 M
 \;.  \label{factor2}
\ee
One can get the vector spectral function by applying pNRQCD calculations as
\be
 \rho^\rmi{pNRQCD}_\rmi{V}(\omega) = 
 \frac12 \Bigl( 1 - e^{-\frac{\omega}{T }}\Bigr)
 \int_{-\infty}^{\infty} \! {\rm d} t \, e^{i \omega t}
 \; C_{>}(t;{\vec{0},\vec{0}})
 \;,  \label{nrqcd_T}
\ee
where $C^{ }_{>}$ is a Wightman function and it is obtained by solving the following 
differential equation 
\be
 \biggl\{ i \partial_t - \biggl[ 2 M 
 + \VT(r) - \frac{\nabla^2_{\vec{r}}}{M}
 \biggr] \biggr\} \, C_{>}^V(t;{\vec{r},\vec{r'}}) = 0 
 \;,  \quad t\neq 0 \;, \label{Seq}
 \ee
 with initial conditions 
 $C_{>}^V(0;{\vec{r},\vec{r'}}) = 6 \Nc\, \delta^{(3)}({\vec{r}-\vec{r'}})\;$. The 
 potential $\VT(r)$ for positive $t$ has the following form \cite{imV,bbr,jacopo}
\be
  \VT(r) = -\alphas C_F \biggl[ \mD^{ } + \frac{\exp(-\mD^{ } r)}{r}
 \biggr] - {i \alphas \CF T } \, \phi(\mD^{ } r)  + \rmO(\alphas^2)\;. \label{expl}
\ee
Here $\phi(x)$ is defined to be
\be
 \phi(x) \equiv 
 2 \int_0^\infty \! \frac{{\rm d} z \, z}{(z^2 +1)^2}
 \biggl[
   1 - \frac{\sin(z x)}{zx} 
 \biggr].
 \label{phi}
\ee
At $\omega<2M$, the spectral function is overestimated. To correct this, we multiply 
$e^{-|\omega-2M|/T}$ to $\phi$. Moreover, at small spatial distances, $r \ll 1/\mD$, 
the thermal potential is replaced by a vacuum one \cite{Burnier:2017bod}.
\begin{figure}[hbt!]
\centering
\includegraphics[width=0.49\textwidth]{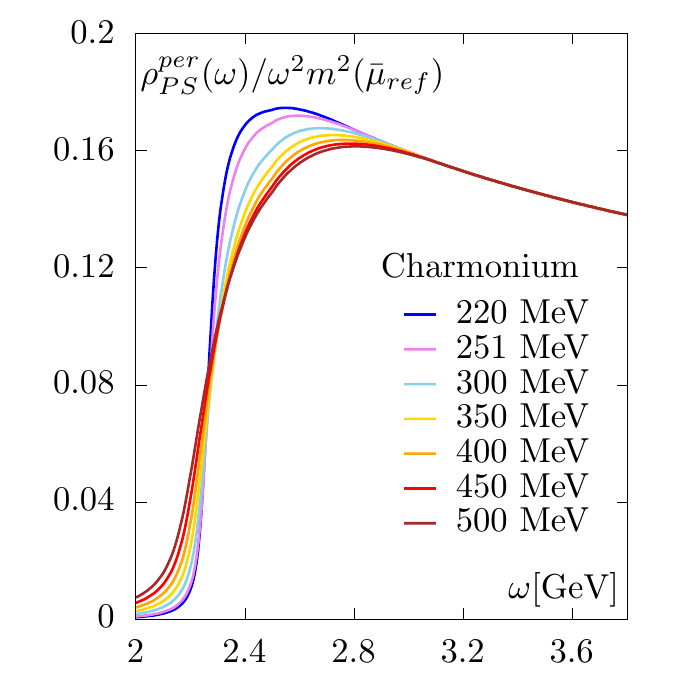}
\includegraphics[width=0.49\textwidth]{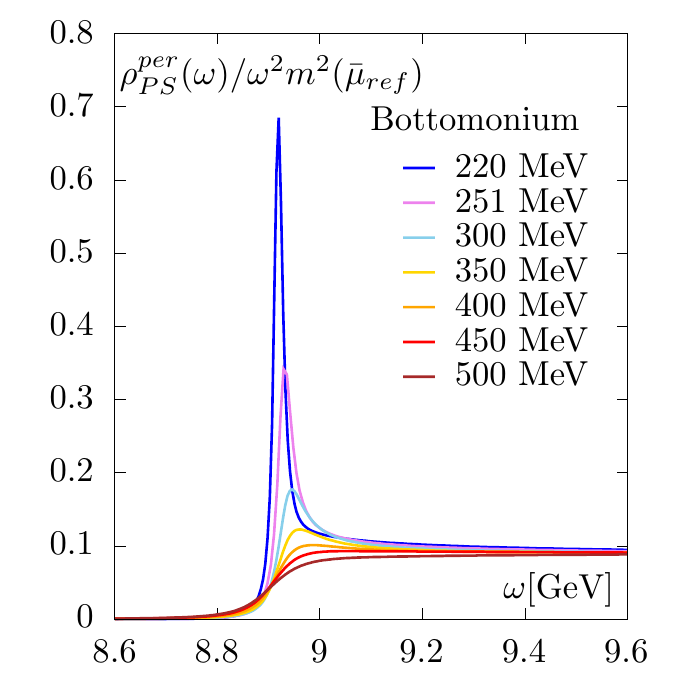}
\caption{Perturbative spectral functions, in the pseudoscalar channel normalised 
by $\omega^2m^2(\bar{\mu}_{ref})$ as a function of $\omega$ for charmonium and bottomonium 
at various temperatures.}\label{fig:purtsp}
\end{figure}
For $\omega>2M$, the spectral function is insensitive to temperature effects, therefore
we also replace it by its vacuum counterpart, which we calculate following  \cite{Burnier:2017bod} and references
therein. The vacuum spectral function, normalised by $\omega^2 m^2(\bar{\mu})$, in
the pseudoscalar channel reads 
\be
 \left. \frac{ \rho^{ }_\rmi{PS}(\omega) }{\omega^2 m^2(\bar{\mu})} \right|^\rmi{vac} 
 \; \equiv \; 
 \frac{\Nc}{8\pi}\, \tilde{R}_\rmi{c}^{p}(\omega) 
 \;. \label{def_v}
\ee
$R$-function is defined as
\ba
 \tilde{R}_\rmi{c}^{p}(\omega) & = &  \tilde{R}^{p(0)}_{ }(\omega) + 
 \frac{\alphas(\bmu) }{\pi} \, \CF \tilde{R}^{p(1)}_{ }(\omega) 
 \nn 
 & + &  
 \biggl( \frac{\alphas(\bmu)}{\pi} \biggr)^2
 \Bigl[ 
   \CF^2\, \tilde{R}^{p(2)}_{A}(\omega)
  + \CF \CA\, \tilde{R}^{p(2)}_{N\!A}(\omega)
  + \CF \TF \Nf\, \tilde{R}^{p(2)}_{l}(\omega)
 \Bigr]  
 + \rmO(\alphas^3)
 \;, \hspace*{6mm} \label{Rp}
\ea
where $\TF \equiv 1/2$, $\CF \equiv (\Nc^2-1)/(2\Nc)$ and
\ba
 \tilde{R}^{p(0)}_{ } & \stackrel{\omega \gg m^{ }(\bmu) }{\approx} & 
  1 
 \;, \label{tR0} \\
 \tilde{R}^{p(1)}_{ } & \stackrel{\omega \gg m^{ }(\bmu) }{\approx} & 
   \frac32 \ln \biggl( \frac{\bmu^2}{\omega^2} \biggr) + \frac{17}4
 \;, \\ 
 \tilde{R}^{p(2)}_{A} & \stackrel{\omega \gg m^{ }(\bmu) }{\approx} & 
   \frac98 \ln^2 \biggl( \frac{\bmu^2}{\omega^2} \biggr) 
   + \frac{105}{16} \ln \biggl( \frac{\bmu^2}{\omega^2} \biggr)
    +\frac{691}{64} - \frac{ 9 \zeta^{ }_2}{4}  
   - \frac{9\zeta^{ }_3}{4}
 \;, \\
 \tilde{R}^{p(2)}_{N\!A} & \stackrel{\omega \gg m^{ }(\bmu) }{\approx} & 
    \frac{11}{16} \ln^2 \biggl( \frac{\bmu^2}{\omega^2} \biggr) 
   + \frac{71}{12} \ln \biggl( \frac{\bmu^2}{\omega^2} \biggr)
   + \frac{893}{64} - \frac{11 \zeta^{ }_2}{8} 
   - \frac{31\zeta^{ }_3}{8}
 \;, \\
 \tilde{R}^{p(2)}_{l} & \stackrel{\omega \gg m^{ }(\bmu) }{\approx} & 
   - \frac{1}{4} \ln^2 \biggl( \frac{\bmu^2}{\omega^2} \biggr) 
   -  \frac{11}{6} \ln \biggl( \frac{\bmu^2}{\omega^2} \biggr)
   - \frac{65}{16} + \frac{\zeta^{ }_2}{2} 
   + \zeta^{ }_3
 \;. \label{tR2l}
\ea
After computing both thermal and vacuum contributions to the spectral function, we
combine them as
\begin{align}
\label{eq-pertspf}
\begin{split}
\rho_\rmi{PS}^\rmi{pert}(\omega)= A^\rmi{match}\rho^{\rmi{pNRQCD}}_\rmi{PS}(\omega)\theta(\omega^\rmi{match}-\omega)
+\rho^\rmi{vac}_\rmi{PS}(\omega)\theta(\omega-\omega^\rmi{match})\;.
\end{split}
\end{align}
In \eq(\ref{eq-pertspf}) we introduce a multiplicative factor $A^\rmi{match}$ to the 
thermal part and it is determined such that both thermal and vacuum parts of the 
spectral function are connected smoothly at some $\omega=\omega^\rmi{match}$, where $2M<\omega^\rmi{match}<3M$. 
The resulting pseudoscalar spectral functions at various temperatures for both
charmonium and bottomonium are shown in \fig(\ref{fig:purtsp}). Unlike charmonium, 
the bottomoniun has a resonance peak that is visible at temperatures below 300 MeV, 
and above 300 MeV bottomonium starts melting.
\section{Spectral reconstruction}
\begin{figure}[hbt!]
\centering
\includegraphics[width=0.49\textwidth]{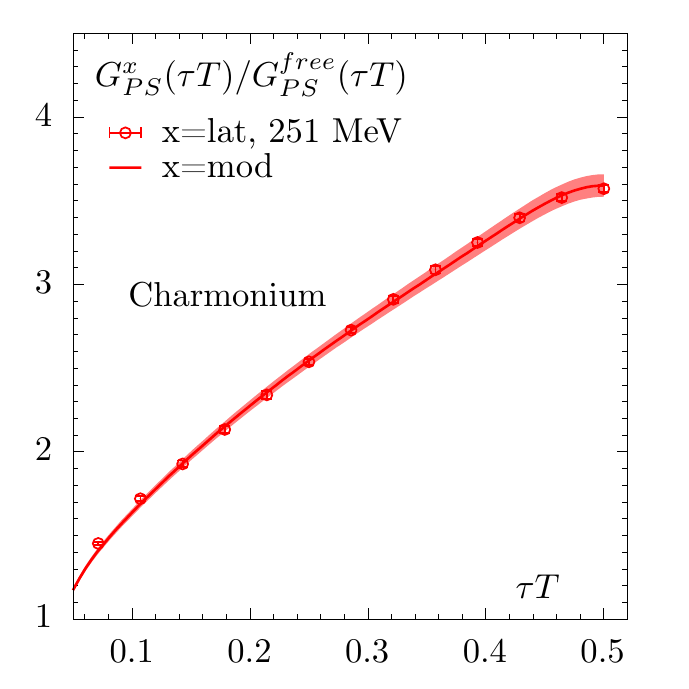}
\includegraphics[width=0.49\textwidth]{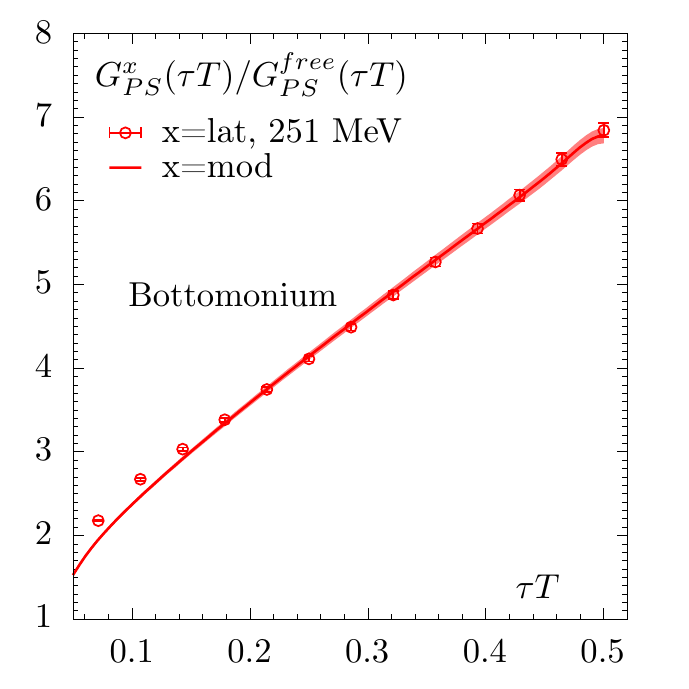}
\caption{Data points represent the original lattice correlator whereas lines 
represent model correlator obtained from \eq(\ref{eq-corr-spec}) for charmonium (left) 
and bottomonium. 
(right).}\label{fig:modsp}
\end{figure}

In this section we will discuss how to model the perturbative spectral functions and 
obtain reconstructed spectral functions. To accommodate the normalization of the 
correlator data and a possible thermal mass shift, we introduce two additional 
parameters $A$ and $B$ to the perturbative spectral function showed in 
\fig(\ref{fig:purtsp}). The model spectral function reads
\be
\rho^\rmi{mod}_\rmi{PS}(\omega)=A\rho^\rmi{pert}_\rmi{PS}(\omega-B)\;.
\ee
This model spectral function is fitted to the unrenormalized pseudoscalar lattice 
correlator that is shown in 
\fig(\ref{fig-corr}), at $T$=251 MeV to determine the parameters $A$ and $B$. Fitting 
the renormalized correlator may change the value of $A$, but $B$ will remain unchanged.
Preliminary results for the model spectral function $\rho^\rmi{mod}$ and original 
correlator data are shown in \fig(\ref{fig:modsp}).
%
\section{Conclusion and Outlook}
We have presented preliminary results of pseudoscalar charmonium and bottomonium correlation functions 
obtained from clover-improved Wilson valence quarks on $2+1$-flavor HISQ gauge field configurations with physical strange quark masses and light quark masses with 
$m_l=m_s/5$ corresponding to $m_\pi$=315 MeV at a temperature of $T$=251 MeV. The tuning of the quark masses was  
done at zero temperature by comparing the spin-averaged quarkonium mass with their corresponding 
experimental values taken from the particle data group. We have constructed the 
perturbative spectral function in the pseudoscalar channel by smoothly matching the 
thermal and vacuum parts. A model is formulated by introducing two parameters to the 
perturbative spectral function. The values of the parameters are determined by fitting 
the model to the lattice correlator data. 
The model describes the lattice 
data reasonably well at large time-slice separations. However, at a small distance, 
there is a discrepancy which might be due to cut-off effects because our measurements 
are still at finite lattice spacing. To study the cut-off effects we will add additional lattice spacings and finally perform a 
continuum extrapolation of the correlation functions, which is work in progress.
This will extend the previous study in the quenched approximation \cite{Burnier:2017bod} towards full QCD.
We also plan to add more temperatures to study the in-medium modifications of quarkonium states in 
the relevant temperature region for heavy ion collision experiments. 
The dynamical light quark degrees of freedom 
are still unphysical in this study. 
Although the dependence of light degrees of freedom on the heavy quark sector may be small, 
to check this dependence we plan to add calculations at 
physical light quark masses. 
This study will also be extended to the vector channel, i.e. unquenching the results \cite{Ding:2021ise} on the in-medium modification of charmonium and bottomonium vector mesons and charm and bottom diffusion coefficients.
\section{ACKNOWLEDGMENTS}
We thank Mikko Laine for helpful and elucidating discussions. We also thank Luis 
Altenkort for the production of gauge configurations and the implementation of meson 
measurements in the QUDA code. This work is supported by the Deutsche 
Forschungsgemeinschaft (DFG, German Research Foundation)-Project number 315477589-TRR 
211. 
The computations in this work were performed on the GPU cluster at Bielefeld University using \texttt{SIMULATeQCD} suite \cite{Bollweg:2021cvl,Mazur:2021zgi}
and QUDA \cite{Clark:2009wm}.
We thank the Bielefeld HPC.NRW team for their support.
%

\end{document}